\documentclass[aps,showpacs,twocolumn,amsmath,amssymb]{revtex4}
\usepackage{graphicx}
\usepackage{dcolumn}
\usepackage{bm}
\begin{document}

\title{ Research on the recurrence relations for the spin-weighted spheroidal harmonics   }
\author{Guihua Tian$^{1,2}$\ \ Wang Hui-hui}\email{tgh-2000@263.net, wendy8151@163.com}  
 \affiliation{1.School of Science, Beijing
University of Posts And Telecommunications. Beijing 100876, China.}
 \affiliation{2.State Key Laboratory of Information Photonics and Optical
Communications, \\ Beijing University of Posts And Telecommunications.
 Beijing 100876, China.}
\date{}
\begin{abstract}
The spin-weighted spheroidal harmonics (SWSHs) are important for the study of the perturbation of the Kerr blackhole, which is relevant with the most tests in relativity. SWSHs with the spin-weight $s=0$ are the spheroidal functions and are easier to study than the counterparts of the spin-weight $s\ne 0$
Through the  methods in super-symmetric quantum mechanics, we give the recurrence relations for SWSHs with different spin weight. These relations  can be  applied to derive SWSHs with $s=\pm 1, \ \pm2 $  from the spheroidal functions.. They also give SWSHs of  $s=-\frac12, \pm \frac32$ from that of $s=\frac12$. These recurrence relations are  first investiged and are very important both in theoretical background and  the astrophysical applications.
\texttt{Keywords}: spin-weighted spheroidal harmonics , recurrence relation, super-symmetric quantum mechanics, shape-invariance
\end{abstract}
\maketitle

\section{Introduction}

The spin-weighted spheroidal harmonics were\begin{eqnarray}
&&\frac{1}{\sin
\theta}\frac{dS}{d\theta}\bigg(\sin \theta \frac{d}{d
\theta}\bigg)+ \bigg[s+a^{2}\omega ^2\cos ^2 \theta \nonumber\\ &&+ 
2a\omega  s\cos \theta -\frac{\left(m+s\cos \theta\right)^{2}}{\sin
^2 \theta}+E\bigg]S =0, \label{angle}
\end{eqnarray}

and 
 first defined by Teukolsky in the context of the perturbation
 to the Kerr black holes.  
where $a$ represent the angular momentum per unit mass of
the rotating black hole and $\omega$ is the frequency of the perturbation fields. 
 The parameter $s$, the
spin-weight of the perturbation fields, could be $s=0, \pm\frac12,
\pm1, \pm2$, and $\Psi$ corresponds to the scalar, neutrino,
electromagnetic or gravitational perturbations respectively.

For simplicity, we will put the parameters $a\omega=\beta$ in the following. Eq.(\ref{angle}) is also called spin-weighted
spheroidal wave equation. Together with the boundary condition  of $S(\theta)$ finite
at $\theta=0,\ \pi$ , Eq.(\ref{angle}) constituted a kind of
the singular Sturm-Liuvelle eigenvalue problem. The eigenfunctions $S_n(\theta), {n=0,1...}$
to the Sturm-Liuvelle problem are called the spin-weighted spheroidal harmonics (SWSHs).
Actually, $S_n(\theta)$ also depends on the parameters $m$ and $s$, and should be denoted as
$S_n(\theta,m,s)$\footnote{SWSHs are denoted as ${}_sS_{lm\omega}(\theta)$ in some references, where $l=n+m$ }. Note we will denote the parameters $m$, $s$ in $S_n$ through their appearance
in the super-potentials $W$, and see the following for details.

SWSHs could help to deepen our understanding of many astrophysical processes modeled as stable problems in Kerr black hole (BH) \cite{1}-\cite{Casals}. 
With   real frequency, they are used to separate the angular dependence of the gravitational radiation produced by perturbation to the Kerr BH.
As angular basis, SWSHs  in Kerr BH are indispensable in all physical problems connecting with the perturbation. A variety of physical situations need to use SWSHs, including the astrophysical problems involving the study of quasi-normal modes (QNMs) of Kerr black hole, quantum field theory in curved space-time and studies of D-branes, etc \cite{7}-\cite{Casals}. For instance, an important application of SWSHs concerning an astrophysical problem  is the determination of black hole parameters from gravitational wave observations, like to determine $M, a$, the source location and the black
hole's spin orientation from the observed waveform. An
investigation of all these issues require the calculation of "scalar products"
 between different quasinormal modes, and,
in particular, between the SWSHs describing their angular
dependence. 

SWSHs are also necessary for  computing the characteristic resonances of Kerr black holes, which will involve the complex parameter.  
In the framework of semi-classical general relativity, it has been conjectured that the highly damped resonances may shed light on the quantum properties of black holes\cite{hod1}-\cite{hod4}. For rotating black holes, these highly damped resonances are characterized by the imaginary part of the frequency approaching infinity. Therefore, the solutions (SWSHs) are important in the theoretical background and   have attracted considerable attentions all the time\cite{1}-\cite{Casals}. Further detailed study on SWSHs is still  very
important.

We have investigated SWSH equations in low frequency cases by the use of the super-symmetric quantum mechanics methods (SUSYQM) and obtained the SWSHs by their recurrence relations \cite{13}-\cite{15}. 

Usually, the spheroidal harmonics (as the $s=0$ case for SWSHs) were thoroughly investigated in past \cite{flamer}. SWSHs for $s\neq0$ only appeared after 1970s and   Eq.(\ref{angle}) is no longer invariant under the  transformation of $\theta \rightarrow\pi-\theta$ whenever $s\neq0$, which might be the cause of the more complex calculation involved
in Eq.(\ref{angle}) and result in the lack of uniform conclusion concerning SWSHs's WKB
approximation\cite{3},\cite{5}-\cite{Casals}. All of these stimulate us to seek the solutions of
Eq.(\ref{angle}) for the case $s\neq0$ in an alternative way, which has been used in the study
of spherical harmonics.

From Eq.(\ref{angle}), it is easy to see that SWSHs reduce to the spin-weighted spherical
harmonics when $\beta=0$ and $S_n(\theta,m,s)e^{im\phi}$ are generally written as ${}_s{Y_{l,m}}$ with $l=m+n$. Generally, the spherical Harmonics $_0{Y_{l,m}}=P_l^m(\cos(\theta))e^{im\phi}$ is in the contents of college level and are easier to grasp, while ${}_s{Y_{l,m}}$ for $s\ne 0$ are not so easy. Nevertheless, one could obtain ${}_s{Y_{l,m}}$ through recurrence relations
as \cite{5}
\begin{eqnarray}
 {}_{s-1}Y_{lm}&=&A_{lm} [\frac{d}{d\theta } + \frac{m + scos\theta }{\sin \theta }]{}_sY_{lm} ,\label{Ym}  \\
_{s+1}Y_{lm}&=&B_{lm}  [\frac{d }{{d\theta }} - \frac{m + scos\theta }{\sin \theta }]
{}_sY_{lm} \label{Ymplus},\\
 A_{lm}&=& \left[(l + s)(l-s+1)\right]^{-\frac12},\nonumber \\  
  B_{lm}&=&\left[(l - s)(l+s+1)\right]^{-\frac12},\nonumber\\ 
  m&=&0,1,2,\dots, l=m,m+1,\cdots \nonumber.
\end{eqnarray}
So, there also are the two ways to obtain ${}_sY_{lm}$, either by solving the corresponding differential equations or utilizing the recurrence relations Eqs.(\ref{Ym})-(\ref{Ymplus}) and ${}_0Y_{lm}=P_l^m(\cos\theta)$.
Actually,
the recurrence relations of Eqs.(\ref{Ym})-(\ref{Ymplus}) are very important in many situation . In the flat space-time, they will provide
a method to obtain the electromagnetic  field contents from the scalar field. 

Similarly, the extension
of Eqs.(\ref{Ym})-(\ref{Ymplus}) to SWSHs is the same important as that in flat case, and will make one obtain
the physical insight of electromagnetic and gravitational perturbation to Kerr black hole from
the information of the scalar perturbation field. So they are worthy of efforts to study and are
the main topics in the paper. The rest of this paper is divided into five sections. After
introduction and review of the SUSYQM and spin-weighted spheroidal harmonics in section 2 and 3.
We re-derive the recurrence relations for the spin-weighted spherical harmonics
 (SWSHs on the condition of $\beta=0$) in section 4. In section 5, we extend the study of section 4 to the spin-weighted spheroidal harmonics with $\beta\neq0$.
 and some conclusion will be given in the final section.

\section{The brief introduction of the super-symmetric quantum mechanics}
  In this section, we give a brief introduction of SUSYQM, which is powerful in solving the Schr\"{o}dinger  equation \cite{16}:
\begin{eqnarray}
-\frac{d^2\psi}{d\theta^2}+[V^-(x)-E]\psi=0 \label{new eq0}.
\end{eqnarray}
A remark on the terms is perhaps necessary. In this paper, we will use the language of SUSYQM, of which the same nomenclatures are used mainly from their mathematical similarity with a Schr\"{o}dinger  equation in quantum mechanics (QM). We make a few remarks: Through Eq.(\ref{new eq0}) has the Schr\"{o}dinger  form, it need not to represent a real QM. Nevertheless, we will apply the nomenclatures in QM, such as the potential energy, the ground energy and state (sometimes ground eigen-vualue and eigenfunction), the excited energies and states, etc in the following for the sake of convenience.

  In SUSYQM, it is mainly to factorize the Hamiltonian $H^-$ of Eq.(\ref{new eq0}) 
  \begin{eqnarray}
{H^-} &=& -\frac{d^2}{dx^2}+V^-(x)+E_0\label{Hamiltonian0},
\end{eqnarray}
as\begin{eqnarray}
{H^-} &=& {{\cal A}^ {\dagger} }{\cal A}^{-} \label{Hamiltonian1},
\end{eqnarray}
where operators ${{\cal A}^ {\dagger} },\  {\cal A}^{-}$ are defined by the super-potential $W$
as \begin{eqnarray}
 {\cal A}^{-}(x)&=& \frac{d}{dx} + W(x), \  \ {\cal A}^{\dagger}(x)= -\frac{d}{dx} + W(x).\label{operatorA-}
 \end{eqnarray}
  Eqs.(\ref{Hamiltonian0})-(\ref{operatorA-}) give the relations of the potential $V$ and the super potential $W$ as
  \begin{eqnarray}
 {W^2}(x) - W'(x) = V^-(x)+E_0\label{potential relations-},
\end{eqnarray}
  which shows the superpotential $W$ is completely determined by the potential $V^-$ and $E_0$, where $E_0 $ is the ground energy of Eq.(\ref{new eq0}). The superpotential $W$ is focus in SUSYQM, it is connected with the ground function $\psi_0$ of Eq.(\ref{new eq0}) by
  \begin{eqnarray}
W(x) &=& - \frac{\psi' _0(x)}{\psi _0(x)}, or
\\
\psi_0&=&N\exp\bigg[-\int Wdx\bigg]\label{psi}.
\end{eqnarray} where $'$ represent the derivative to $x$.
Interchange the order of the operators ${{\cal A}^ {\dagger} },\  {\cal A}^{-}$  in Eq.(\ref{Hamiltonian1}) will give the partner Hamiltonian $H^+$ of $H^-$:
 \begin{eqnarray}
{H^+} &=& {\cal A}^ {-} {\cal A}^{\dagger} =  - \frac{d^2}{dx^2} + V^+\nonumber\\
&=&  - \frac{d^2}{dx^2} + {W^2}(x) + W'(x) \label{Hamiltonian2},
\end{eqnarray}
where the super-potential $W$ is connected with the potential $V^+$
\begin{eqnarray}
 {V^ + }(x)&=&{W^2}(x) + W'(x) .\label{potential and super relation2}
\end{eqnarray}
 The partner Hamiltonians $H^-,\ H^+$ shares the same  eigen-energies except the ground energy $E_0$ \cite{16}, and their eigenfunctions $\psi_n^-,\ \psi_n^+$ , that is 
 \begin{eqnarray}
H^{-}\psi_n^{-}= (E_n-E_0)\psi_n^{-},\ n=0, 1,2,\ \cdots
\\
H^{+}\psi_n^{+}= (E_n-E_0)\psi_n^{+},\ n=1,2,\ \cdots
\end{eqnarray} 
  are related by 
  \begin{eqnarray}
\psi_n^{+}= {\cal A}^ {-} \psi_n^{-},\  \
\psi_n^{-}= {\cal A}^{\dagger} \psi_n^{+}.
\end{eqnarray}

The ground eigenfunction of $H^-$ is given directly by the super-potential $W$.
However, the excited state functions or eigen-functions cannot be obtained directly by the super-potential $W$,
 and needs some other properties of $W$, the shape invariance property. In order to
 introduce the shape-invariance concept, the super-potential $W$ will be denoted by $W(x,a_1)$
with $a_1$ representing some parameters in $W$. So the
pair of SUSY partner potentials $V^{\mp}(x,a_1)$ correspondingly become
\begin{equation}
 V^{\mp}(x,a_1)=W^2(x,a_1)\mp W'(x,a_1)\label{vpm},
\end{equation}
and the corresponding partner Hamiltonians are
\begin{eqnarray}
H_1^{-}(x;a_1)= -\frac{d^2}{dx^2}+ {V^- }(x;a_1)= {\cal A}^ {\dagger} (x;a_1){\cal A}^{-}(x;a_1), \label{Hamiltonian13}\\
H_1^{+}(x;a_1)= -\frac{d^2}{dx^2} + {V^+ }(x;a_1)= {\cal A}^{-}(x;a_1){\cal A}^ {\dagger} (x;a_1).\label{Hamiltonian4}
\end{eqnarray}
If the pair of partner potentials $V^{\pm}$  are similar in shape
and different only from parameters,  that is
\begin{equation} \label{shape invariant} V^+(x;a_1) = V^-(x;a_2)+R(a_1),
\end{equation}
where $a_2=f(a_1)$ and the remainder $R(a_1)$ is independent of
$x$, then the super-potential $W$ is said to be of shape-invariance.
Through $a_2=f(a_1)$ in the super-potential $W(x,a_2)$, one will have the new partner Hamiltonians:
\begin{eqnarray}
H_2^{-}(x;a_2)= -\frac{d^2}{dx^2}+ {V^- }(x;a_2)= {\cal A}^ {\dagger} (x;a_2){\cal A}^{-}(x;a_2), 
\label{Hamiltonian13}\\
H_2^{+}(x;a_2)= -\frac{d^2}{dx^2} + {V^+ }(x;a_2)= {\cal A}^{-}(x;a_2){\cal A}^ {\dagger} (x;a_2), \label{Hamiltonian4}
\end{eqnarray}
The shape-invariance properties will result in the relations of the eigen-energies and eigen-functions of $H_2^{-}(x;a_2),\  H_1^{+}(x;a_1)$ and make the Hamiltonian $H_1^{-}(x;a_1)$
 completely integrable \cite{16}. That is, 
 \begin{eqnarray}
 H_1^{-}(x;a_1)= -\frac{d^2}{dx^2}+ {V^- }(x;a_1)= {\cal A}^ {\dagger} (x;a_1){\cal A}^{-}(x;a_1), \label{Hamiltonian13}\\
 H_2^{-}(x;a_2)= -\frac{d^2}{dx^2} + {V^- }(x;a_2)= {\cal A}^ {\dagger} (x;a_2){\cal A}^{-}(x;a_2)\nonumber\\
 = H_1^{+}(x;a_1)-R(a_1)= {\cal A}^{-}(x;a_1) {\cal A}^ {\dagger} (x;a_1)-R(a_1).\label{Hamiltonian4}
\end{eqnarray}
meets the relation
 \begin{eqnarray}
 H_2^{-}(x;a_2)&=& {\cal A}^ {\dagger} (x;a_2){\cal A}^{-} (x;a_2)\nonumber\\ =  H_1^{+}(x;a_1)-R(a_1)&=&  {\cal A}^{-}(x;a_1) {\cal A}^ {\dagger} (x;a_1)-R(a_1).\label{Hamiltonian41}
\end{eqnarray}

with their wave functions $\psi_n^{-}(x;a_1), \psi_n^{-}(x;a_2)$ satisfying respectively
\begin{eqnarray}
H_1^{-}(x;a_1)\psi_n^{-}(x;a_1)= E_n^1\psi_n^{-}(x;a_1),\\
H_2^{-}(x;a_2)\psi_n^{-}(x;a_2)= E_n^2\psi_n^{-}(x;a_2).
\end{eqnarray}
$\psi_n^{-}(x;a_1)$,$\psi_n^{-}(x;a_2)$ can be connected by the pair of operator ${\cal A}^{-}(x;a_1)$,
${\cal A}^{\dagger}(x;a_1)$ as \cite{16}
\begin{eqnarray}
\psi_n^{-}(x;a_1)= {\cal A}^{\dagger}(x;a_1)\psi_n^{-}(x;a_2)\label{relation1}\\
\psi_n^{-}(x;a_2)= {\cal A}^{-}(x;a_1)\psi_n^{-}(x;a_1)\label{relation2}
\end{eqnarray}
and $E_n^1=E_n^2+R(a_1)$. The Hamiltonians $H^-(x,a_1)$, $H^-(x,a_2)$ are  the hierarchy of Hamiltonians we construct. Similarly, further Hamiltonian  can be built by the further shape-invariance of the super potential $W(x,a_2)$, as we will use in the following.   It is obvious that Eq.(\ref{relation1}) and Eq.(\ref{relation2}) are the recurrence relations, which will be applied to study SWSHs.

\section{Review of the spin-weighted spheroidal harmonics}

In order to apply  SUSYQM to Eq.(\ref{angle}), one should use the transformation \cite{13}-\cite{10}
\begin{eqnarray}
S(\theta)=\frac{\psi(\theta)}{\sqrt{sin\theta}}.\label{transfer}
\end{eqnarray}
to make it into a Schr\"{o}dinger form 
\begin{eqnarray}
&&\frac{d^2\psi}{d\theta^2}+\bigg[\frac{1}{4}+s +\beta^{2}\cos ^2 \theta
-2s\beta \cos\theta\nonumber\\
&& -\frac{(m+s\cos\theta)^2-\frac{1}{4}}{\sin
^2\theta}+E\bigg]\psi=0.\label{new eq1}
\end{eqnarray}
The super-potential $W$ is expanded as the series sum of
the parameter $\beta$ as
\begin{equation}
W=W_0+\sum_{n=1}^{\infty}\beta^nW_{n}.\label{super-potential expansion}
\end{equation}
In Refs.\cite{13}-\cite{18}, the authors have studied the Eq.(\ref{new eq1}) with the general formula for $W_n$:
\begin{equation}
W_{n}(\theta)=\sum_{k=1}^{[\frac{n}{2}]}a_{n,k}\sin^{2k-1}\theta\cos\theta+\sum_{k=1}^{[\frac{n+1}{2}]}b_{n,k}\sin^{2k-1}\theta.\label{w_n}
\end{equation}
The first several expressions of $W_n$ have  been obtained in Refs.\cite{17}-\cite{15}, of which we give the first three ones as follows:
\begin{eqnarray}
{W_0}(\theta )&=&\frac{{b_{0,0} + a_{0,0}\cos \theta}}{{\sin\theta}}=-\frac{{s+(m+\frac{1}{2})\cos \theta }}{{\sin \theta }},\label{w00}\\
W_{1}(\theta)&=&b_{1,1}\sin\theta=-\frac{s}{m+1}\sin\theta,\label{w11}\\
W_2(\theta)&=&b_{2,1}sin\theta+a_{2,1}sin\theta\cos\theta,\nonumber\\
&=&-\frac{(m+s+1)(m-s+1)s}{(m+1)^3(2m+3)}sin\theta,\nonumber\\
&+&\frac{(m+s+1)(m-s+1)}{(m+1)^2(2m+3)}sin\theta\cos\theta.\label{w22}
\end{eqnarray}
The rest coefficients $a_{n,j},b_{n,j}$ of $W_n$ are described in detail in  Refs.\cite{17}-\cite{15}.

\section{The recurrence relations for the spin-weighted spherical harmonics}
As previously stated, SWSHs become to the spin-weighted spherical harmonics under special condition of $\beta=0$.
In this section, we study the recurrence relations about the different spin-s fields for the spin-weighted spherical harmonics by the methods in SUSYQM, 
and verify that the results thus obtained are consistent with   Eq.(\ref{Ym}) previously obtained in Ref.\cite{5}.

We construct the idea of the hierarchy of Hamiltonians\cite{16}:
\[H = {H_1},\,{H_2},\,{H_3},\, \ldots \]
to study the kind of recurrence relations with the same parameter $m$ but different $s$.
here we choose  the first three Hamiltonians $ {H_1}, {H_2}, {H_3}$ to obtain the relations among them.
\subsection{The useful formulas from the shape-invariance of the potentials}

The first three Hamiltonians $ {H_1}, {H_2}, {H_3}$ corresponding to the Hamiltonian in section 2 are rewritten
as $H_1(\theta;a_1), H_2(\theta;a_2), H_3(\theta;a_3)$, whose corresponding potentials have the shape invariance property. In above, we put
 $a_1=(A_0,B_0),  a_2=(C_0,D_0),  a_3=(E_0,F_0)$, where $A_0, B_0, C_0, D_0, E_0, F_0$ are  constants.

Due to $\beta=0$, the super potential becomes $W=W_0$. The super potentials corresponding to the first three Hamiltonian $H_1(\theta;A_0,B_0),H_2(\theta;C_0,D_0),H_3(\theta;E_0,F_0)$ are expressed by
\begin{eqnarray}
&&{W_0}(\theta,A_0,B_0 ) =  \frac{{B_0 + A_0\cos \theta }}{{\sin \theta }},\\
&&{W_0}(\theta,C_0,D_0 ) =  \frac{{D_0 + C_0\cos \theta }}{{\sin \theta }},\\
&&{W_0}(\theta,E_0,F_0 ) =  \frac{{F_0 + E_0\cos \theta }}{{\sin \theta }}.
\end{eqnarray}
The definition of the shape-invariance of the potential gives the following 
\begin{eqnarray}
V^+_1(\theta;A_{0},B_{0}) &=& V^-_1(\theta;C_{0},D_{0})+R(A_{0},B_{0}),\\
V^+_2(\theta; C_{0},D_{0})&=& V^-_2(\theta; E_{0},F_{0})+R(C_{0},D_{0}).
\end{eqnarray}
 The above equations will provide  the relations
 among the undetermined constants $A_0, B_0, C_0, D_0, E_0, F_0$ , which turn out  to be four ones corresponding to the four cases.
 The first is
 \begin{eqnarray}
C_0={A_0}-1,\ \
D_0={B_0},\label{relation da1}\\
E_0={C_0}-1=A_0-2,\ \
F_0={D_0}=B_0,\label{relation fb1}
\end{eqnarray}
and  has been applied  to obtain the spin-weighted spheroidal functions $S_{n}(\theta,m,s),\ n>m$ from $S_{n}(\theta,m,s),\ n=m$ in \cite{18}-\cite{19}.
The second one is 
\begin{eqnarray}
C_0=-{A_0},\ \
D_0=-{B_0},\label{relation da2}\\
E_0=-{C_0}=A_0,\label{relation ea2}\\
F_0=-{D_0}=B_0,\label{relation fb2}
\end{eqnarray}
and is trivial one.
The last two are 
\begin{eqnarray}
C_0=-{B_0}-\frac{1}{2},\ \
D_0=-{A_0}+\frac{1}{2},\label{relation da3}\\
E_0=-{D_0}-\frac{1}{2}=A_0-1,\label{relation ea3}\\
F_0=-{C_0}+\frac{1}{2}=B_0+1,\label{relation fb3}
\end{eqnarray}
and 
\begin{eqnarray}
C_0={B_0}-\frac{1}{2},\ \
D_0={A_0}-\frac{1}{2},\label{relation da4}\\
E_0={D_0}-\frac{1}{2}=A_0-1,\label{relation ea4}\\
F_0={C_0}+\frac{1}{2}=B_0-1,\label{relation fb4}
\end{eqnarray}
which will be used to obtain the recurrence relations in the following.
Some contents need to be noted here. The relations  (\ref{relation da3})-(\ref{relation fb4}) are different from that in Refs.\cite{18}-\cite{19}, 
and are important
for the study of the recurrence relations of spin-weighted spheroidal harmonics with different spin $s$.
\subsection{The recurrence relation from $_sY_{l,m+1}$ to $_{s-1}Y_{l,m+1}$}
From $A_0=-(m + \frac{1}{2})$, $B_0=-s$ and Eqs.(\ref{w00}),  (\ref{relation da3})-(\ref{relation fb3}), it is easy to obtain
\begin{eqnarray}
&&{W_0}(\theta;A_0,B_0 ) =- \frac{{s + (m + \frac{1}{2})\cos \theta }}{{\sin \theta }},\label{super potential Wa}\\
&&{W_0}(\theta;C_0,D_0 ) =\frac{{(m+1) + (s- \frac{1}{2})\cos \theta }}{{\sin \theta }},\label{super potential Wb}\\
&&{W_0}(\theta;E_0,F_0 ) =- \frac{{(s-1) + (m + \frac{3}{2})\cos \theta }}{{\sin \theta }}.\label{super potential Wc}
\end{eqnarray}

 To use Eq.(\ref{relation1})-(\ref{relation2}) repeatedly will produce the following recurrence relation
 \begin{equation}
{\psi _n}(\theta;E_0,F_0) = {\cal A}^{-}(\theta;C_0,D_0){\cal A}^{-}(\theta;A_0,B_0){\psi _{n}}(\theta;A_0,B_0),\label{psi-ef}
\end{equation}
Then one  defines
\begin{equation}
{\psi _n}(\theta;\widetilde{C_0},\widetilde{D_0}) = {\cal A}^{-}(\theta;A_0,B_0){\psi _{n}}(\theta;A_0,B_0)\label{connection}
\end{equation}
with $\widetilde{C_0}=-(m+\frac{3}{2}),\widetilde{D_0}=-s$, which have been obtained in Refs.\cite{18}-\cite{19}. The Hamiltonian  for ${\psi _n}(\theta;\widetilde{C_0},\widetilde{D_0})$ is\begin{equation}
H^{-}(\theta;\widetilde{C_0},\widetilde{D_0})= {{\cal A}^ {\dagger} (\theta;\widetilde{C_0},\widetilde{D_0})}{\cal A}^{-}(\theta;\widetilde{C_0},\widetilde{D_0}).
\end{equation}
The spin-weighted spherical harmonics $S(\theta)$  with the condition of spin $s$ and magnetic quantum number $m+1$ are usually  denoted by ${}_s{Y_{l,m+1}}$ . Following this tradition, we see that  Eq.(\ref{transfer}) is
\begin{eqnarray}
{}_{s-1}{Y_{l,m+1}}=\frac{\psi_{n}(\theta;E_0,F_0)}{\sqrt{sin\theta}},\label{transfer21} \\
{}_s{Y_{l,m+1}}=\frac{{\psi_{n}}(\theta;\widetilde{C_0},\widetilde{D_0})}{\sqrt{sin\theta}} \label{transfer22}
\end{eqnarray}
with $l=m+1+n$. So it is easy to observe  that Eq.(\ref{psi-ef}) about the eigenfunctions ${\psi _n}$
is the recurrence relation of ${}_s{Y_{l,m}}$  with the condition of spin $s\rightarrow{s-1}$ and the same parameter $m+1$.

By Eqs.(\ref{operatorA-}), (\ref{super potential Wa})-(\ref{psi-ef}), (\ref{transfer21})-(\ref{transfer22}), we have
\begin{eqnarray}
\bigg[\frac{d }{{d\theta }} + \frac{{(m+1)+ s cos\theta }}{{\sin \theta }}\bigg]{}_s{Y_{l,m+1}} ={}_{s - 1}{Y_{l,m+1}}\label{qm+1},
\end{eqnarray}
which is exactly the recurrence relations of ${}_s{Y_{l,m+1}}\rightarrow {}_{s-1}{Y_{l,m+1}}$. which satisfy the condition of same m but different spin s.

With the parameter $m$ changing  into $m+1$ and ignoring the normalized constant, it is easy to see that
 Eq.(\ref{qm+1}) is of the same form as that in Eq.(\ref{Ym}). Hence the recurrence relation Eq.(\ref{qm+1}) for SWSHs under the condition $\beta=0$
is consistent with recurrence relations for the spin-weighted spherical harmonics from Ref.\cite{5}.

\subsection{The recurrence relation from $_sY_{l,m+1}$ to $_{s+1}Y_{l,m+1}$}
Similarly to the last subsection, we will use the relations   (\ref{relation da4})-(\ref{relation fb4})  
to obtain the recurrence (\ref{Ymplus}).
With \[A_0=-(m +\frac{1}{2}),B_0=-s,\]  we see that 
\begin{eqnarray}C_0&=&-(s+\frac{1}{2}),\ D_0=-(m+1 ),\\
  E_0&=&-(m +1+ \frac{1}{2}),\ F_0=-s-1.\end{eqnarray} 
  The Hamiltonians ${\cal A}^{+}(\theta;A_0,B_0){\cal A}^{-}(\theta;A_0,B_0)$ and ${\cal A}^{+}(\theta;E_0,F_0){\cal A}^{-}(\theta;E_0,F_0)$ correspond to the transformed ones of spin-weighted spherical harmonics with $m, s$ and $m+1, s+1$ respectively.
Define $\psi _{n}(\theta;A_0,B_0),\ n=0,1,\cdots $ are the eigen-functions for the Hamiltonian ${\cal A}^{+}(\theta;A_0,B_0){\cal A}^{-}(\theta;A_0,B_0)$ , then
\begin{equation}
{\psi _n}(\theta;E_0,F_0) = {\cal A}^{-}(\theta;C_0,D_0){\cal A}^{-}(\theta;A_0,B_0){\psi _{n}}(\theta;A_0,B_0),\label{psi-ef22}
\end{equation}
 are the eigenfunctions for or the Hamiltonian ${\cal A}^{+}(\theta;E_0,F_0){\cal A}^{-}(\theta;E_0,F_0)$. So
 \begin{equation}
{}_{s+1}{Y_{l,m+1}}=\frac1{\sqrt{sin\theta}}{\psi_{n}}(\theta;{E_0},{F_0})\label{transfer23}
\end{equation}
with $l=m+1+n$.
With Eqs.(\ref{connection}), (\ref{transfer22}), we have
 \begin{equation}
{}_{s+1}{Y_{l,m+1}}=\frac1{\sqrt{sin\theta}}{\cal A}^{-}(\theta;C_0,D_0)\bigg[\sqrt{sin\theta}  _{s}{Y_{l,m+1}}\bigg]\label{transfer33}
\end{equation}
with $l=m+1+n$.
Eq.(\ref{transfer33}) turns out to be
\begin{equation}
{}_{s+1}{Y_{l,m+1}}=\bigg[\frac{d}{{d\theta }} - \frac{{(m+1)+ s cos\theta }}{{\sin \theta }}\bigg]{}_s{Y_{l,m+1}} \label{2qm+1},
\end{equation}
So we obtain the recurrence relations Eqs.(\ref{qm+1}), (\ref{2qm+1}) for the spin-weighted spherical harmonics,which are consistent with Eqs.(\ref{Ym})-(\ref{Ymplus}), and we will extend the methods to the spin-weighted spheroidal harmonics with $\beta\ne 0$ in the following.

\section{The recurrence relations for the SWSHs under the condition of $\beta\ne 0$  I}

In this part, we study the recurrence relations for SWSHs under the common condition of $\beta\neq0$ by the methods in SUSYQM. There three parts in the section. Part A involves the introduction of the parameters into the super-potential. Part B and C are the extension of Eqs.(\ref{qm+1}), (\ref{2qm+1}) to the case of $\beta\neq 0$.
\subsection{The parameters for introduction of the shape-invariance potential}
In order to  make use of the shape-invariance
properties of the hierarchy of Hamiltonians in SUSYQM, in Ref.\cite{18}
the authors introduce some constant parameters $A_{i,j}=1,\ B_{i,j}=1$ into
the general formula of super-potential $W$ . For the sake of simplicity, we denote the collection of $A_{i,j},B_{i,j}, \ i\le \frac n2+1, j\le n$ by $\mathcal{A}_{n},\mathcal{B}_{n}$ and all the collection of $\mathcal{A}_{n},\mathcal{B}_{n},\ n=0,\ n=1, \cdots $ by $\mathcal{A},\mathcal{B}$
\begin{eqnarray}
W(\theta; \mathcal{A},\mathcal{B})=W_0(\theta; \mathcal{A}_{0},\mathcal{B}_{0})+\sum_{n=1}^{\infty}\beta^nW_n(\theta; \mathcal{A}_{n},\mathcal{B}_{n}),
\end{eqnarray}
where
\begin{eqnarray}
W_0(\theta;  \mathcal{A}_{0},\mathcal{B}_{0})=\frac{-A_{0,0}(m+\frac12)\cos
\theta-sB_{0,0}}{\sin\theta},
\end{eqnarray}
\begin{eqnarray}
W_n(\theta; \mathcal{A}_{n},\mathcal{B}_{n})&=& \sum_{j=1}^{[\frac{n+1}{2}]}\bar{b}_{n,j}\sin^{2j-1}\theta \nonumber \\
&+&\cos \theta\sum_{j=1}^{[\frac{n}{2}]}\bar{a}_{n,j}\sin^{2j-1}\theta\label{super-W},
\end{eqnarray}
with
\begin{eqnarray}
\bar{a}_{n,j} =A_{n,j}a_{n,j},\  \bar{b}_{n,j} =B_{n,j}b_{n,j}.
\end{eqnarray}
Some of $a_{n,j}, b_{n,j}$ for $n\le 2$ are given in Eqs.(\ref{w00})-(\ref{w22}). For the other terms of   $a_{n,j}, b_{n,j}$, see Refs.\cite{17}-\cite{15}.
By the same way, We also represent the collection $C_{i,j},D_{i,j}, \frac n2+1, j\le n$  by  $\mathcal{C}_n, \mathcal{D}_n$ and $E_{i,j},F_{i,j}, \frac n2+1, j\le n$  by  $\mathcal{E}_n, \mathcal{F}_n$ . Introducing  parameters $\mathcal{C}, \mathcal{D}$ to be the set of $\mathcal{C}_{n},\mathcal{D}_{n}, \ n=0,1,2,\cdots$  and  $\mathcal{E}, \mathcal{F}$ to be the set of $\mathcal{E}_{n},\mathcal{F}_{n}, \ n=0,1,2,\cdots$.

The super potential can be written as  $W(\theta; \mathcal{C}, \mathcal{D})$. Then 
\begin{eqnarray}
&&W(\theta; \mathcal{C},\mathcal{D})=W_0(\theta; C_{0,0},D_{0,0})\nonumber \\
&&+\sum_{n=1}^{\infty}\beta^nW_n(\theta; \mathcal{C}_{n},\mathcal{D}_{n}),
\end{eqnarray}
with all the forms of $W(\theta; \mathcal{C},\mathcal{D})$ being the same as that of $W(\theta; \mathcal{A},\mathcal{B})$ . Similarly we define the super potential $W(\theta; \mathcal{E},\mathcal{F})$ as above.

Then, $V^{\pm}(\theta; \mathcal{A},\mathcal{B})$ are defined as
\begin{eqnarray}
V^{\pm}(\theta; \mathcal{A},\mathcal{B})&=&W^2(\theta; \mathcal{A},\mathcal{B})\pm
W'(\theta; \mathcal{A},\mathcal{B})\nonumber\\&=&\sum_{n=0}^{\infty}\beta^nV^{\pm}_n(\theta; \mathcal{A}_n,\mathcal{B}_n),\\
V^{\pm}(\theta; \mathcal{C},\mathcal{D})&=&W^2(\theta; \mathcal{C},\mathcal{D})\pm
W'(\theta; \mathcal{C},\mathcal{D})\nonumber\\&=&\sum_{n=0}^{\infty}\beta^nV^{\pm}_n(\theta; \mathcal{C}_n,\mathcal{D}_n),\\
V^{\pm}(\theta; \mathcal{E},\mathcal{F})&=&W^2(\theta; \mathcal{E},\mathcal{F})\pm
W'(\theta; \mathcal{E},\mathcal{F})\nonumber\\&=&\sum_{n=0}^{\infty}\beta^nV^{\pm}_n(\theta; \mathcal{E}_n,\mathcal{F}_n)
\end{eqnarray}
The  shape-invariance properties require for all $n\ge 0$
\begin{eqnarray}
V^{+}_n(\theta; \mathcal{A}_{n},\mathcal{B}_{n})=V^{-}_n(\theta; \mathcal{C}_{n},\mathcal{D}_{n})+R_{n;m}(\mathcal{A}_{n},\mathcal{B}_{n})\label{shape-invariance v-pm},\\
V^{+}_n(\theta; \mathcal{C}_{n},\mathcal{D}_{n})=V^{-}_n(\theta; \mathcal{E}_{n},\mathcal{F}_{n})+R_{n;m}(\mathcal{C}_{n},\mathcal{D}_{n})\label{shape-invariance v-pm2}
\end{eqnarray}  
with  $R_{n;m}(\mathcal{A}_{n},\mathcal{B}_{n}), R_{n;m}(\mathcal{C}_{n},\mathcal{D}_{n})$ pure quantities. All parameters $\mathcal{C}, \mathcal{D}, \mathcal{E}, \mathcal{F}$ can be derived from the parameters $\mathcal{A}, \mathcal{B}$, which all are equal to one. 

\subsection{The recurrence relations obtained from relations (\ref{relation da3})-(\ref{relation fb3})}
In order to extend the formula (\ref{qm+1}) to 
SWSHs with $\beta\ne 0$, one applies  Eqs.(\ref{relation da3})-(\ref{relation fb3}) to obtain 
\begin{eqnarray}
C_{0,0}&=&\frac{{-2sB_{0,0}+1}}{2m+1},\\
D_{0,0}&=&-\frac{(2m+1)A_{0,0}+1}{2s} \\
E_{0,0}&=&\frac{-2sD_{0,0}+1}{2m+1},\\
F_{0,0}&=&-\frac{(2m+1)C_{0,0}+1}{2s},
\end{eqnarray}
which also can be derived from Eqs.(\ref{shape-invariance v-pm})-(\ref{shape-invariance v-pm2}) under the condition $n=0$. When $n=1$, Eqs.(\ref{shape-invariance v-pm})-(\ref{shape-invariance v-pm2}) gives
 \begin{eqnarray}
D_{1,1}&=&\frac{(2m+1){A_{0,0}}-1}{(2m+1){C_{0,0}}+1}B_{1,1},\label{d11 and b110}\\
F_{1,1}&=&\frac{(2m+1){C_{0,0}}-1}{(2m+1){E_{0,0}}+1}D_{1,1},\label{d11 and b11}
\end{eqnarray}
which become
\begin{eqnarray}
D_{1,1}&=&-\frac{(2m+1){A_{0,0}}-1}{2sB_{0,0}+2}{B_{1,1}},\\
F_{1,1}&=&\frac{sB_{0,0}[(2m+1)A_{0,0}-1]}{[(2m+1)A_{0,0}+3][sB_{0,0}+1]}B_{1,1}.\label{d11 and b11 2}
\end{eqnarray}
For the general form of $n\ge2$, similar calculation could proceed to give the corresponding 
$\mathcal{C}_{n},\mathcal{D}_n$ and $\mathcal{E}_{n},\mathcal{F}_n$, see appendix for details. Here we just show the results as following:
\begin{eqnarray}
D_{n,p}&=&-2s\frac{D_{0,0}a_{n,p}}{\alpha_{p}b_{n,p}}C_{n,p}-\frac{U_{n,p}}{\alpha_{p}b_{n,p}}\label{D-np}\\
C_{n,p-1}&=&
\frac{\bigg(\alpha_{p}-\frac{4s^2D^2_{0,0}}{\alpha_{p}}\bigg)a_{n,p}}{(\alpha_{p}-1)a_{n,p-1}}C_{n,p}\nonumber\\ &+&
\frac{\check{U}_{n,p}-\frac{2sD_{0,0}}{\alpha_{p}}U_{n,p}}
{(\alpha_{p}-1)a_{n,p-1}}, \ p=2,3,\ldots, [\frac{n+2}2],\label{c-np}
\end{eqnarray}
with $D_{n,[\frac{n+1}2]+1}=C_{n,[\frac{n}2]+1}=0$, $\alpha_{p}=(2m+1)C_{0,0}+(2p-1)$ and $U_{n,p}, \check{U}_{n,p}$ being given by Eqs.(\ref{bigunp})-(\ref{cbigunp}) in the appendix.

 Similarly, the quantities $E_{n,j},F_{n,j}$ also can be calculated through $\mathcal{C}_{n},\mathcal{D}_{n}$ by:
\begin{eqnarray}
F_{n,p}&=&-2s\frac{{{F_{0,0}}{a_{n,p}}}}{{{\gamma _p}{b_{n,p}}}}{E_{n,p}}-\frac{{{Y_{n,p}}}}{{{\gamma _p}{b_{n,p}}}}\label{F-np}\\
E_{n,p-1}&=&2s\frac{F_{0,0}b_{n,p}}{({\gamma _p}-1)a_{n,p-1}}{F_{n,p}}+\frac{{{\gamma _p}{a_{n,p}}}}{{({\gamma _p} - 1){a_{n,p - 1}}}}{E_{n,p}}\nonumber\\
&+& \frac{\check{Y}_{n,p}}{{({\gamma _p} - 1){a_{n,p - 1}}}}\label{E-np}
\end{eqnarray}
 where $Y_{n,p}, \check{Y}_{n,p}, \gamma_p$ are given by Eqs.(\ref{bigynp})-(\ref{cbigynp}), (\ref{ccbigvnp}) in the appendix. 
 
 The eigenfunctions $\psi_n$ are obtained by the recurrence relation from Eq.(\ref{psi-ef}):
\begin{eqnarray}
&& \psi_n(\theta;\widetilde{\mathcal{C}},\widetilde{\mathcal{D}})={\cal A}^{-}(\theta;\mathcal{A},\mathcal{B}){\psi_{n}}(\theta;\mathcal{A},\mathcal{B}),\\
&&{\psi _n}(\theta;\mathcal{E},\mathcal{F})= {\cal A}^{-}(\theta;\mathcal{C},\mathcal{D}) \psi_n(\theta;\widetilde{\mathcal{C}},\widetilde{\mathcal{D}})\label{psi-ef3}\\
&&{\cal A}^{-}(\theta;\mathcal{A},\mathcal{B})=\frac d{d\theta}+W(\theta;\mathcal{A},\mathcal{B})\\
&&{\cal A}^{-}(\theta;\mathcal{C},\mathcal{D})=\frac d{d\theta}+W(\theta;\mathcal{C},\mathcal{D}).\label{last two recurrence3}
\end{eqnarray}

Finally, through transferring eigenfunctions $\psi_n$ into the spin-weighted spheroidal harmonics ${S _n}$ by means of Eq.(\ref{transfer}), that is 
\begin{eqnarray}
S_n(\theta;\mathcal{A},\mathcal{B})=\frac{\psi_n(\theta;\mathcal{A},\mathcal{B})}{\sqrt{sin\theta}},\label{transfer31}\\
S_{n}(\theta;\widetilde{\mathcal{C}},\widetilde{\mathcal{D}})=\frac{\psi_n(\theta;\widetilde{\mathcal{C}},\widetilde{\mathcal{D}})}{\sqrt{sin\theta}},\label{transfer3}\\
S_n(\theta;\mathcal{E},\mathcal{F})=\frac{\psi_n(\theta;\mathcal{E},\mathcal{F})}{\sqrt{sin\theta}},\label{transfer34}
\end{eqnarray}
we will rewrite $S_n$ by the tradition as $S_n(\theta,m,s)$, that is, 
\begin{eqnarray}
{S _n}(\theta;\mathcal{A},\mathcal{B})&=&S_n(\theta,m,s),\label{psi-ab}\\
S_{n}(\theta;\widetilde{\mathcal{C}},\widetilde{\mathcal{D}})&=&S_n(\theta,m+1,s),\label{psi-cd}\\
{S _n}(\theta;\mathcal{E},\mathcal{F})&=&S_n(\theta,m+1,s-1).
\end{eqnarray}
So Eq.(\ref{psi-ef3}) becomes
\begin{eqnarray}
&&S_n(\theta,m+1,s-1)\nonumber\\
&=&
[\frac d{d\theta}+\frac{{\cos \theta }}{{2\sin \theta }}+W(\theta;\mathcal{C},\mathcal{D})]S_n(\theta,m+1,s)\nonumber\\
&=&\bigg[\frac d{d\theta}+\frac{s\cos \theta +(m+1)}{\sin \theta }+\beta \frac{ms}{(m+1)(s+1)}\sin\theta \nonumber\\
&+&\sum_{n=2}^{\infty}\beta^nW_n(\theta;\mathcal{C}_n,\mathcal{D}_n)\bigg]S_n(\theta,m+1,s).\label{last-relations11}
\end{eqnarray}

\subsection{The recurrence relations obtained from relations (\ref{relation da4})-(\ref{relation fb4})}

We could proceed as before to obtain the extension of recurrence relations (\ref{Ymplus}) for the spin-weighted spheroidal harmonics from Eqs. (\ref{relation da4})-(\ref{relation fb4}), which 
tell us 
\begin{eqnarray}
C_{0,0}&=&\frac{2sB_{0,0}+1}{{2m+1}},\\
D_{0,0}&=&\frac{(2m+1)A_{0,0}+1}{2s} ,\\
E_{0,0}&=&\frac{2sD_{0,0}+1}{{2m+1}},\\
F_{0,0}&=&\frac{(2m+1)C_{0,0}+1}{2s} .
\end{eqnarray}
 Eqs.(\ref{shape-invariance v-pm})-(\ref{shape-invariance v-pm2}) under the condition $n=1$ give the same formula as that of Eqs.(\ref{d11 and b11} )
\begin{eqnarray}
D_{1,1}&=&\frac{(2m+1){A_{0,0}}-1}{(2m+1){C_{0,0}}+1}B_{1,1},\\
F_{1,1}&=&\frac{(2m+1){C_{0,0}}-1}{(2m+1){E_{0,0}}+1}D_{1,1}\label{d11 and b11 4}.
\end{eqnarray}
So $D_{1,1},\ F_{1,1}$ turn out as
\begin{eqnarray}
D_{1,1}&=&\frac{{(2m+1){C_{0,0}}-1}}{{2s{D_{0,0}}+2}}{D_{1,1}}
\nonumber\\
&=&\frac{(2m+1){A_{0,0}}-1}{2sB_{0,0}+2}{B_{1,1}},\\
F_{1,1}&=&\frac{sB_{0,0}[(2m+1)A_{0,0}-1]}{[(2m+1)A_{0,0}+3][sB_{0,0}+1]}B_{1,1}.\label{d11 and b11 3}
\end{eqnarray}
$D_{1,1}$ is different from Eqs.(\ref{d11 and b11 2}).
As the formula Eqs.(\ref{d11 and b110})-(\ref{d11 and b11}) for $D_{1,1},\ F_{1,1}$ are the same as Eqs.(\ref{d11 and b11 4}) , so we can use the formulas in the last subsection to obtain the subsequent $\mathcal{C}_n,\ \mathcal{D}_n,\ \mathcal{E}_n,\ \mathcal{F}_n$ for $n\ge 2$ as Eqs.(\ref{D-np})-(\ref{E-np}). Note that actually the dependence of $\mathcal{C}_n,\ \mathcal{D}_n,\ \mathcal{E}_n,\ \mathcal{F}_n$ on the parameters $\mathcal{A},\ \mathcal{B}$ is different from that in the last subsection.
With Eqs.(\ref{psi-ab})-(\ref{psi-cd}) and the definition of
\begin{eqnarray}
S_n(\theta,m+1,s+1)=S _n(\theta;\mathcal{E},\mathcal{F})=\frac{\psi_n(\theta;\mathcal{E},\mathcal{F})}{\sqrt{sin\theta}},\label{transfer4}
\end{eqnarray}
the recurrence relations (\ref{psi-ef3}) now become
\begin{eqnarray}
&&S_n(\theta,m+1,s+1)\nonumber\\
&=&
[\frac d{d\theta}+\frac{{\cos \theta }}{{2\sin \theta }}+W(\theta;\mathcal{C},\mathcal{D})]S_n(\theta,m+1,s)\\
&=&\bigg[\frac d{d\theta}- \frac{s\cos \theta +(m+1)}{\sin \theta }-\beta \frac{ms}{(m+1)(s+1)}\sin\theta \nonumber\\
&+&\sum_{n=2}^{\infty}\beta^nW_n(\theta;\mathcal{C}_n,\mathcal{D}_n)\bigg]S_n(\theta,m+1,s).
\ \label{last-relations22}
\end{eqnarray}
The above equations are just the extension of formula (\ref{Ymplus}) to the spin-weighted spheroidal harmonics.

\section{The recurrence relations for the SWSHs under the condition of $\beta\neq0$ II}
In study of SWSHs with the method of SYSUQM, there are two kind forms for the super-potential $W$. The first one is that of (\ref{super-potential expansion}), Eq.(\ref{w_n}) \cite{10}, and is applied thus far to obtain the recurrence relations for different SWSHs \cite{13}-\cite{19} and in the above section. The second one is the form \cite{20}-\cite{15}:
\begin{eqnarray}
W&=&W_0+\sum_{n=1}^{\infty}\beta^nW_{n}\label{super-potential expansion2}\\
W_{n}(\theta)&=&\sin\theta \sum_{k=0}^{n-1}a_{n,k}\cos^{k}\theta.\label{w_n2}
\end{eqnarray}
except for $a_{0,0}$,  We will write $a_{i,j} $ for all $i, j$ and admit $a_{i,j} =0$ whenever one of the conditions $i\ge 1$ and $0\le j\le i-1$ is violated. This will simplifying the calculation involved later.
Please note that these parameters $a_{n,k}$ in the two forms of the super-potentials generally  represent different quantities except for the case $n\le1$. For the same of simplicity, we do not denote new forms to them. The actually quantities of $a_{n,k}$ are given in Ref.\cite{15}.
The current section will provide the recurrence relations similar to Eqs.(\ref{last-relations11}), (\ref{last-relations22}) in the second form of the super-potential. Let $\mathcal{A}_n$ denote the set $B_{0,0}, A_{i,j}, i\le n$ , and   $\mathcal{C}_n$ the set $D_{0,0}, C_{i,j}, i\le n$ and $\mathcal{E}_n$ the set $F_{0,0}, E_{i,j}, i\le n$.  Similarly $\mathcal{A}, \mathcal{C}$ represent the same physical contents as in the above section 
\begin{eqnarray}
W(\theta; \mathcal{A})&=&W_0(A_{0,0}, B_{0,0})+\sum_{n=1}^{\infty}\beta^nW_{n}(\theta, \mathcal{A}_n)\label{super-potential expansion21}\\
W_{n}(\theta; \mathcal{A}_n)&=&\sin\theta \sum_{k=0}^{n-1}a_{n,k}\mathcal{A}_{n,k}\cos^{k}\theta,\label{w_n21}
\end{eqnarray}
and
\begin{eqnarray}
W(\theta; \mathcal{C})&=&W_0(C_{0,0}, D_{0,0})+\sum_{n=1}^{\infty}\beta^nW_{n}(\theta, \mathcal{C}_n)\label{super-potential expansion31}\\
W_{n}(\theta; \mathcal{C}_n)&=&\sin\theta \sum_{k=0}^{n-1}a_{n,k}\mathcal{C}_{n,k}\cos^{k}\theta.\label{w_n32}
\end{eqnarray}
and
\begin{eqnarray}
W(\theta; \mathcal{E})&=&W_0(E_{0,0}, F_{0,0})+\sum_{n=1}^{\infty}\beta^nW_{n}(\theta, \mathcal{E}_n)\label{super-potential expansion32}\\
W_{n}(\theta; \mathcal{E}_n)&=&\sin\theta \sum_{k=0}^{n-1}a_{n,k}\mathcal{E}_{n,k}\cos^{k}\theta.\label{w_n33}
\end{eqnarray}
the partner potentials related with the three super potentials are
\begin{eqnarray}
V^{\pm}(\theta; \mathcal{A})&=&W^2(\theta; \mathcal{A})\pm
W'(\theta; \mathcal{A})=\sum_{n=0}^{\infty}\beta^nV^{\pm}_n(\theta; \mathcal{A}),\\
V^{\pm}(\theta; \mathcal{C})&=&W^2(\theta; \mathcal{C})\pm
W'(\theta; \mathcal{C})=\sum_{n=0}^{\infty}\beta^nV^{\pm}_n(\theta; \mathcal{C}),\\
V^{\pm}(\theta; \mathcal{E})&=&W^2(\theta; \mathcal{E})\pm
W'(\theta; \mathcal{E})=\sum_{n=0}^{\infty}\beta^nV^{\pm}_n(\theta; \mathcal{E})
\end{eqnarray}
and the shape-invariance properties require the following to be met, that is
\begin{eqnarray}
V^{+}_n(\theta; \mathcal{A}_{n})=V^{-}_n(\theta; \mathcal{C}_{n})+R_{n;m}(\mathcal{A}_{n})\label{shape-invariance v-pmA},\\
V^{+}_n(\theta; \mathcal{C}_{n})=V^{-}_n(\theta; \mathcal{E}_{n})+R_{n;m}(\mathcal{C}_{n})\label{shape-invariance v-pmC}
\end{eqnarray}  
with  $R_{n;m}(\mathcal{A}_{n}), R_{n;m}(\mathcal{C}_{n})$ pure quantities. All parameters $\mathcal{C}, \mathcal{E}$ can be derived from the parameters $\mathcal{A}$, which all are equal to one. 
Here we give the following formulae for $V^{\pm}_n(\theta; \mathcal{A}_{n}), n\ge2$ for later use. 

\begin{eqnarray}
&&V^{\pm}_n(\theta; \mathcal{A}_{n})=-2\bar{b}_{0,0}\bar{a}_{n,0}\pm \bar{a}_{n,1}+\sum_{k=0}^{n-1}\bar{a}_{n-k,0}\bar{a}_{k,0}\nonumber\\
&+&
\sum_{k=0}^{n-1}\bigg[\sum_{l=0}^{n-1}\sum_{i=0}^{n-1}(\bar{a}_{n-l,k+1-i}-\bar{a}_{n-l,k-i} )\bar{a}_{l,i}\nonumber\\
&&\ \ \ \ \ \ \ \ +2\bar{b}_{0,0}\bar{a}_{n,k+1}\mp (k+2)\bar{a}_{n,k+2}\nonumber\\
&& \ \ \ \ \ \ \ \ +(2\bar{a}_{0,0}\pm (k+1))\bar{a}_{n,k}
\bigg]\cos^{k+1}\theta,
\end{eqnarray}
where $\bar{b}_{0,0}=b_{0,0}B_{0,0}, \bar{a}_{n,k}=a_{n,k}A_{n,k}$.
Similarly one could provide $V^{\pm}_n(\theta; \mathcal{C}_{n}), V^{\pm}_n(\theta; \mathcal{E}_{n}), n\ge 2$, which will be omitted.

In order to extend (\ref{Ym}) to SWSHs, we first utilize the relations (\ref{relation da3})-(\ref{relation fb3}) to obtain the parameters $\mathcal{C}, \mathcal{E}$ for fulfill the requirement of the shape-invariance properties of the super-potential.
As stated before, $W_n, n\le1$ are the same in both Eqs.(\ref{super-potential expansion})-(\ref{w_n}) and  Eqs.({super-potential expansion2})-(\ref{w_n2}), so the results in subsection VA calculated from the shape-invariance requirements are valid for obtaining the parameters $\mathcal{C}_1$, and will be just rewritten here
 as
\begin{eqnarray}
C_{0,0}&=&\frac{{-2sB_{0,0}+1}}{2m+1},\\
D_{0,0}&=&-\frac{(2m+1)A_{0,0}+1}{2s} \\
E_{0,0}&=&\frac{-2sD_{0,0}+1}{2m+1},\\
F_{0,0}&=&-\frac{(2m+1)C_{0,0}+1}{2s}\\
D_{1,1}&=&-\frac{(2m+1){A_{0,0}}-1}{2sB_{0,0}+2}{B_{1,1}},\\
F_{1,1}&=&\frac{sB_{0,0}[(2m+1)A_{0,0}-1]}{[(2m+1)A_{0,0}+3][sB_{0,0}+1]}B_{1,1}.
\end{eqnarray}
For $n\ge2 $, we can derive 
\begin{eqnarray}
C_{n,k}&=&-\frac{2b_{0,0}D_{0,0}a_{n,k+1}C_{n,k+1}+ (k+2)a_{n,k+2}C_{n,k+2}}{(2a_{0,0}C_{0,0}- (k+1)a_{n,k}}\nonumber\\
&+& \frac{X_{n,k}}{2a_{0,0}C_{0,0}- (k+1)a_{n,k}}
\end{eqnarray}
\begin{eqnarray}
E_{n,k}&=&-\frac{2b_{0,0}F_{0,0}a_{n,k+1}E_{n,k+1}+ (k+2)a_{n,k+2}E_{n,k+2}}{(2a_{0,0}E_{0,0}- (k+1))a_{n,k}}\nonumber\\
&+& \frac{\check{X}_{n,k}}{(2a_{0,0}E_{0,0}- (k+1))a_{n,k}}
\end{eqnarray}
where
$X_{n,k}, \check{X}_{n,k}$ are 
\begin{eqnarray}
X_{n,k}
&=&
\sum_{i,l=0}^{n-1}\bigg(a_{n-l,p}A_{n-l,p}-a_{n-l,p-1}A_{n-l,p-1} \bigg)a_{l,i}A_{l,i}\nonumber\\
&+&
2b_{0,0}B_{0,0}a_{n,k+1}A_{n,k+1}- (k+2)a_{n,k+2}A_{n,k+2}\nonumber\\
&+&
\bigg(2a_{0,0}A_{0,0}+ (k+1)\bigg)a_{n,k}A_{n,k}
\end{eqnarray}
and 
\begin{eqnarray}
\check{X}_{n,k}
&=&
\sum_{i,l=0}^{n-1}\bigg(a_{n-l,p}C_{n-l,p}-a_{n-l,p-1}C_{n-l,p-1} \bigg)a_{l,i}C_{l,i}\nonumber\\
&+&
2b_{0,0}D_{0,0}a_{n,k+1}C_{n,k+1}- (k+2)a_{n,k+2}C_{n,k+2}\nonumber\\
&+&
\bigg(2a_{0,0}C_{0,0}+ (k+1)\bigg)a_{n,k}C_{n,k}
\end{eqnarray}
where $p=k+1-i$. Therefore, the extended recurrence relations could be written as before
\begin{eqnarray}
&&S_n(\theta,m+1,s-1)\nonumber\\
&=&
[\frac d{d\theta}+\frac{{\cos \theta }}{{2\sin \theta }}+W(\theta;\mathcal{C})]S_n(\theta,m+1,s)\nonumber\\
&=&\bigg[\frac d{d\theta}+\frac{s\cos \theta +(m+1)}{\sin \theta }+\beta\frac{ms}{(m+1)(s+1)}\sin\theta \nonumber\\
&+&\sum_{n=2}^{\infty}\beta^nW_n(\theta;\mathcal{C}_n)\bigg]S_n(\theta,m+1,s).\label{last-relations33}
\end{eqnarray}
in the same way, we also extend (\ref{relation da4})-\ref{relation fb4}) as
\begin{eqnarray}
&&S_n(\theta,m+1,s+1)\nonumber\\
&=&
[\frac d{d\theta}+\frac{{\cos \theta }}{{2\sin \theta }}+W(\theta;\mathcal{C})]S_n(\theta,m+1,s)\nonumber\\
&=&\bigg[\frac d{d\theta}-\frac{s\cos \theta +(m+1)}{\sin \theta }-\beta\frac{ms}{(m+1)(s+1)}\sin\theta \nonumber\\
&+&\sum_{n=2}^{\infty}\beta^nW_n(\theta;\mathcal{C}_n)\bigg]S_n(\theta,m+1,s).\label{last-relations44}
\end{eqnarray}

\section{Discussion and Conclusion }
In summary, we have obtained the recurrence relations for the spin-weight spheroidal harmonics with the different spins, which are consistent in the case of $\beta=0$ with the result given by R. Breuer et al. in  Ref.\cite{5}. Our methods apply SUSYQM to SWSHs, where the super-potential $W$ is the key concept. By the sue of super-potential as in Eq.(\ref{w_n}), we obtain the recurrence relations Eqs.(\ref{last-relations11}), (\ref{last-relations22}). Similarly, we give recurrence relations (\ref{last-relations33})-(\ref{last-relations44}) through Eq.(\ref{w_n21}). Of course, these two kinds relations should be the same, as the first three terms shows. Whether using Eqs.(\ref{last-relations11}), (\ref{last-relations22}) or (\ref{last-relations33})-(\ref{last-relations44})  depends on the form of SWSHs. By the methods of SUSYQM, we have investigated SWSHs thoroughly and the results will surely can be utilized numerically to obtain detailed informations on SWSHs.

However, our study give no information about the recurrence relations from half integer to integer spins of SWSHs. Further study should be how to extend this relations to that of SWSHs of spins being half integer and integer.

We give some examples about the present paper application.
The results of Eqs.(\ref{last-relations11}), (\ref{last-relations22}) make one obtain the spin-weighted spheroidal harmonics $S_n(\theta,m,s-1)$ and $S_n(\theta,m,s+1)$ just from the spheroidal harmonics $S_n(\theta,m,s)|_{s=0}$. 
The spirit of this kind manufacturing process could also be used to study the radial Teukolsky equation.
Thus, one will obtain the properties of the perturbation field $\psi$ about $s=1$ and $s=2$ through
the scalar perturbation field by use of recurrence relations, and this will give us new insight for the Kerr black hole perturbation study.
Also the recurrence relations  provide some information about the normalization
constants concerning SWSHs, as it has already been done in Ref.\cite{6}. Further extension of the study in the paper might be the application of the methods to study the radial Teukolsky equations, which might provide a new view to the stable problem of the Kerr black hole.

\acknowledgments
The work was partly supported by the National Natural Science of China (No. 10875018)
and the Major State Basic Research Development Program of China (973 Program: No.2010CB923202).

\section{Appendix: Detailed Calculation }
\begin{widetext}
For the general form of $n\ge0$, one simplifies the expressions of
$V^{ \pm}_n(\mathcal{A}_{n},\mathcal{B}_{n})$. With the help of
\begin{eqnarray}
&&W^2(\theta; \mathcal{A}_{n},\mathcal{B}_{n})=W_0^2+\sum_{n=1}^{\infty}\beta^nW_n(\theta; \mathcal{A}_{n},\mathcal{B}_{n})+
\sum_{n=2}^{\infty}\beta^n\sum_{k=1}^{n-1}W_k(\theta; \mathcal{A}_{k},\mathcal{B}_{k})W_{n-k}(\theta; \mathcal{A}_{n-k},\mathcal{B}_{n-k}),\\
&&W'(\theta; \mathcal{A}_{n},\mathcal{B}_{n})=W'_0(\theta; A_{0,0},B_{0,0})+\sum_{n=1}^{\infty}\beta^nW'_n(\theta; \mathcal{A}_{n},\mathcal{B}_{n}),
\end{eqnarray}
by the use of Eq.(\ref{w_n}), we obtain the formulae for $V_n^{\pm}$ in the case $n\ge 1$ as following
\begin{eqnarray}
V^{\pm}_n(\theta; \mathcal{A}_{n},\mathcal{B}_{n})&=&\cos\theta\sum_{p=1}^{[\frac{n+1}{2}]}\bigg[P^{\pm}_{n,p}(\mathcal{A}_{n},\mathcal{B}_{n})+G_{n,p}(A_{n-1},B_{n-1})\bigg]\sin^{2p-2}\theta \nonumber\\
&&
+\sum_{p=1}^{[\frac{n}{2}]+1}\bigg[Q^{\pm}_{n,p}(\mathcal{A}_{n},\mathcal{B}_{n})+H_{n,p}(A_{n-1},B_{n-1})\bigg]\sin^{2p-2}\theta
\end{eqnarray}
where
\begin{eqnarray}
P^{\pm}_{n,p}(\mathcal{A}_{n},\mathcal{B}_{n})&=& 2b_{0,0}B_{0,0}A_{n,p}a_{n,p}+\bigg(2a_{0,0}A_{0,0}\pm (2p-1)\bigg)B_{n,p}b_{n,p},\label{P-np}\\
Q^{\pm}_{n,p}(\mathcal{A}_{n},\mathcal{B}_{n})&=&2b_{0,0}B_{0,0}B_{n,p} b_{n,p}+2a_{0,0}A_{0,0}a_{n,p} A_{n,p}-2a_{0,0}A_{0,0}a_{n,p-1} A_{n,p-1}\nonumber\\
&&\pm (2p-1)A_{n,p}a_{n,p}\mp (2p-2)A_{n,p-1}a_{n,p-1}\label{Q-np}\\
G_{n,p}(\mathcal{A}_{n-1},\mathcal{B}_{n-1})&=&\sum_{k=1}^{n-1}\sum_{j=1}^{[\frac {n}2]+1}
\bigg[b_{k,p-j}B_{k,p-j}a_{n-k,j}A_{n-k,j}+a_{k,p-j}A_{k,p-j}b_{n-k,j}B_{n-k,j}\bigg]\label{b-p2},\\
H_{n,p}(\mathcal{A}_{n-1},\mathcal{B}_{n-1})&=&\sum_{k=1}^{n-1}\sum_{j=1}^{[\frac {n}2]+1}
\bigg[b_{k,p-j}B_{k,p-j}b_{n-k,j}B_{n-k,j} \nonumber\\
&&
+a_{k,p-j}A_{k,p-j}a_{n-k,j}A_{n-k,j}-a_{k,p-1-j}A_{k,p-1-j}a_{n-k,j}A_{n-k,j}\bigg]\label{b-p3},
\end{eqnarray}
where  $a_{n,j} =0$ whenever $j < 0$ or $j >[\frac {n}2]$ and $b_{n,j}= 0$ whenever$j < 0$ or $j >[\frac {n+1}2]$. Similarly, we have
\begin{eqnarray}
V^{\pm}_n(\theta; \mathcal{C}_{n},\mathcal{D}_{n})&=&\cos\theta\sum_{p=1}^{[\frac{n+1}{2}]}\bigg[P^{\pm}_{n,p}(\mathcal{C}_{n},\mathcal{D}_{n})+G_{n,p}(C_{n-1},D_{n-1})\bigg]\sin^{2p-2}\theta \nonumber\\
&& +\sum_{p=1}^{[\frac{n}{2}]+1}\bigg[Q^{\pm}_{n,p}(\mathcal{C}_{n},\mathcal{D}_{n})+H_{n,p}(C_{n-1},D_{n-1})\bigg]\sin^{2p-2}\theta
\end{eqnarray}
where
\begin{eqnarray}
P^{\pm}_{n,p}(\mathcal{C}_{n},\mathcal{D}_{n})&=& 2b_{0,0}D_{0,0}C_{n,p}a_{n,p}+\bigg(2a_{0,0}C_{0,0}\pm (2p-1)\bigg)D_{n,p}b_{n,p},\label{P-np2}\\
Q^{\pm}_{n,p}(\mathcal{C}_{n},\mathcal{D}_{n})&=&2b_{0,0}D_{0,0}b_{n,p} B_{n,p}+2a_{0,0}C_{0,0}a_{n,p} C_{n,p}-2a_{0,0}C_{0,0}a_{n,p-1} C_{n,p-1}\nonumber\\
&&\pm (2p-1)C_{n,p}a_{n,p}\mp (2p-2)C_{n,p-1}a_{n,p-1}\label{Q-np2}\\
G_{n,p}(\mathcal{C}_{n-1},\mathcal{D}_{n-1})&=&\sum_{k=1}^{n-1}\sum_{j=1}^{[\frac {n}2]+1}
\bigg[b_{k,p-j}D_{k,p-j}a_{n-k,j}C_{n-k,j}+a_{k,p-j}C_{k,p-j}b_{n-k,j}D_{n-k,j}\bigg]\label{b-p22},\\
H_{n,p}(\mathcal{C}_{n-1},\mathcal{D}_{n-1})&=&\sum_{k=1}^{n-1}\sum_{j=1}^{[\frac {n}2]+1}
\bigg[b_{k,p-j}D_{k,p-j}b_{n-k,j}D_{n-k,j} \nonumber\\
&&
+a_{k,p-j}C_{k,p-j}a_{n-k,j}C_{n-k,j}-a_{k,p-1-j}C_{k,p-1-j}a_{n-k,j}C_{n-k,j}\bigg]\label{b-p32},
\end{eqnarray}
where  $a_{n,j} =0$ whenever $j < 0$ or $j >[\frac {n}2]$ and $b_{n,j}= 0$ whenever$j < 0$ or $j >[\frac {n+1}2]$ .

One could use the shape-invariance equation Eq.(\ref{shape-invariance v-pm})
to obtain 
\begin{eqnarray}
 P^-_{n,p}(\mathcal{C}_{n},\mathcal{D}_{n})&=&P^+_{n,p}(\mathcal{A}_{n},\mathcal{B}_{n})+G_{n,p}(\mathcal{A}_{n-1},\mathcal{B}_{n-1})-G_{n,p}(\mathcal{C}_{n-1},\mathcal{D}_{n-1})\\
 &\equiv&U_{n,p},\label{bigunp}\\
 Q^-_{n,p}(\mathcal{C}_{n},\mathcal{D}_{n})&=&H_{n,p}(\mathcal{A}_{n-1},\mathcal{B}_{n-1})+Q^+_{n,p}(\mathcal{A}_{n},\mathcal{B}_{n})-H_{n,p}(\mathcal{C}_{n-1},\mathcal{D}_{n-1})\\
 &\equiv&
 \check{U}_{n,p},\label{cbigunp}
\end{eqnarray}
and  from Eq.(\ref{P-np2})and Eq.(\ref{Q-np2}) we can get
\begin{eqnarray}
P^-_{n,p}(\mathcal{C}_{n},\mathcal{D}_{n})&=& -\alpha_{p}D_{n,p}b_{n,p}-D_{0,0}C_{n,p}a_{n,p}\nonumber\\
Q^-_{n,p}(\mathcal{C}_{n},\mathcal{D}_{n})&=&
-\alpha_{p}C_{n,p}a_{n,p}-D_{0,0}D_{n,p}b_{n,p}+
(\alpha_{p}-1)C_{n,p-1}a_{n,p-1}\nonumber,
\end{eqnarray}
where $\alpha_{p}=(2m+1)C_{0,0}+(2p-1)$. In the same way, we have
\begin{eqnarray}
Y_{n,p}&=&P^+_{n,p}(\mathcal{C}_{n},\mathcal{D}_{n})+G_{n,p}(\mathcal{C}_{n-1},\mathcal{D}_{n-1})-G_{n,p}(\mathcal{E}_{n-1},\mathcal{F}_{n-1})\\
 &=&P^-_{n,p}(\mathcal{E}_{n},\mathcal{F}_{n})\label{bigynp}\\
\check{Y}_{n,p}&=&H_{n,p}(\mathcal{C}_{n-1},\mathcal{D}_{n-1})+Q^+_{n,p}(\mathcal{C}_{n},\mathcal{D}_{n})-H_{n,p}(\mathcal{E}_{n-1},\mathcal{F}_{n-1})\\
&=&
 Q^-_{n,p}(E_{n,j},F_{n,j}),\label{cbigynp}
\end{eqnarray}
where 
\begin{eqnarray}
P^-_{n,p}(\mathcal{E}_{n},\mathcal{F}_{n})&=& -\gamma_{p}E_{n,p}a_{n,p}-2sF_{0,0}F_{n,p}b_{n,p}\label{bigvnp}\\
 Q^-_{n,p}(E_{n,j},F_{n,j})&=&
-\gamma_{p}E_{n,p}a_{n,p}-2sF_{0,0}F_{n,p}b_{n,p}+
(\gamma_{p}-1)E_{n,p-1}a_{n,p-1},\label{cbigvnp}\\
\gamma_{p}&=&(2m+1)E_{0,0}+(2p-1)\label{ccbigvnp}.
\end{eqnarray}

\end{widetext}

\end{document}